\documentstyle[prl,aps,twocolumn,epsfig]{revtex}
\begin{document}
\newcommand{\babar}{BABAR}
\twocolumn[\hsize\textwidth\columnwidth\hsize\csname
  @twocolumnfalse\endcsname
%
\draft
\begin{center}
{\hspace{15.0cm}COLO-HEP 478}
\end{center}

\title{Mass Determination Method for
 $\tilde{e}_{L,R}$ Above Production Threshold}
\author{Mihai Dima, James Barron, Anthony Johnson, \\
Luke Hamilton, Uriel Nauenberg, Matthew Route,\\
David Staszak, Matthew Stolte, Tara Turner}
\address{Department of Physics, University of Colorado, Boulder.}
\date{\today}
\maketitle

\begin{abstract}
The determination of the masses of Supersymmetric particles such as
the selectron, for energies above threshold using the energy
end-points method is subject to signal deconvolution difficulties and to 
Standard Model and Supersymmetry backgrounds. The important features of 
$\tilde{e}_{R,L}^\pm$ production are used to design an experimentally robust 
method, with good resolution, both for determining the $\tilde{e}_{R,L}$ and 
the $\tilde{\chi}_1^0$ masses and for suppressing backgrounds. Additional 
features, such as the determination of the relative leptonic branching ratios 
of the selectron are present in the method. 
\end{abstract}
\pacs{Keywords:  
SUSY Particle Production, Experimental Mass Determination, Selectron}
 ] 


The determination of Supersymmetric particle masses using the energy 
end-point method~\cite{bib:tsukamoto}~is well known. The measurement of
selectron masses is subject to two experimental difficulties: on the
one hand the energy distribution of the visible particles in the event 
is an overlap of 4 ``box like'' distributions due to the production channels
$\tilde{e}_R^+\tilde{e}_R^-, \tilde{e}_L^+\tilde{e}_L^-, \tilde{e}_R^+\tilde{e}_L^-, \tilde{e}_L^+\tilde{e}_R^-$ and on the other, the Supersymmetry 
(SUSY) signal is masked by very large Standard Model (SM) backgrounds, such as
W$^+$W$^-$ and $\gamma^*$$\gamma^*$. The consequences are, for the former the 
difficulty of resolving overlapping ``box"-edges in the lower energy region 
making it hard to determine which edge is due to which SUSY particle 
decay, while for the latter the masking of the SUSY-signal by large
SM backgrounds.
 
Many other studies~\cite{bib:Feng1,bib:Feng2,bib:Martyn,bib:Baer,bib:Colorado}
of the determination of supersymmetric masses via the energy spectrum of the
observed particles have been carried out showing the usefulness of
the technique and the levels of accuracy possible. The complications in 
the selectron spectrum are being solved here. 

During Snowmass-2001 we realised that the difference of the observed positron 
and electron (from the $\tilde{e}_{L,R}$ decay) energy distributions enhanced 
by the difference in cross sections with incident electron polarization 
(possible only with a Linear Collider) can solve these problems and provide a 
method by which we can determine the masses unambiguously with good resolution.
In addition it offers new features, built in redundancy, the determination
of the partial leptonic branching ratios to the channels $\tilde{\chi}^\pm
_1\nu_e$ and $\tilde{\chi}^0_1e$. Information about the  $\tilde{\chi}^\pm_1$ 
and $\tilde{\tau}^\pm_1$ masses may, in principle, also be determined.

 The $e^+, e^-$ energy distributions difference
is given by:
\begin{equation}
\Delta(E) = Lumi \cdot
            (\sigma_{RL} - \sigma_{LR}) \cdot
            [R'_{box}(E) - L'_{box}(E)]
\end{equation}
where, for instance,  $\sigma_{LR}$  is the production cross section for 
$\tilde{e}^+_L \tilde{e}^-_R$, while $R'(E)$ and $L'(E)$ are the appropriate 
energy box distributions (for the incident electron right-handed and
left-handed polarization) each normalised to unity. The method presented here 
removes the contribution to the energy spectrum from the reactions producing 
$\tilde{e}_R^+\tilde{e}_R^-$ and $\tilde{e}_L^+\tilde{e}_L^-$, which solves the
mass measurement difficulty and reduces the mass errors.

Asymmetric boosts are present when $\tilde{e}^\pm_L$ and $\tilde{e}^\pm_R$ are 
produced in the same reaction. The values of the boosts being in this case:
 \begin{eqnarray}
\gamma_{\tilde{e}_L} &=& \frac{1}{2M_{\tilde{e}_L}}\sqrt{s} + \frac{M^2_{\tilde{e}_L} - M^2_{\tilde{e}_R}}{2M_{\tilde{e}_L} \sqrt{s}}\cr\cr
\gamma_{\tilde{e}_R} &=& \frac{1}{2M_{\tilde{e}_R}}\sqrt{s} + \frac{M^2_{\tilde{e}_R} - M^2_{\tilde{e}_L}}{2M_{\tilde{e}_R} \sqrt{s}}
 \end{eqnarray}

 The energies of the electron and positron which are the particles visible in 
the event are related to the decay CM energy. Since $m_e \ll M_{L,R}$ we obtain
the lower and higher bounds of the electron/positron energy distribution in the
LAB reference frame:
 \begin{eqnarray}
 E_{e^\pm}(HI) &=& E_{e^\pm}^*(\gamma + \sqrt{\gamma^2 -1}) \cr \cr
 E_{e^\pm}(LO) &=& E_{e^\pm}^*(\gamma - \sqrt{\gamma^2 -1}) =
 E_{e^\pm}^*/(\gamma + \sqrt{\gamma^2 -1})
 \end{eqnarray}
 Solving for $E_{e^\pm}^*$ and $\gamma$ we obtain:
 \begin{eqnarray}
 E_{e^\pm}^* &=& \sqrt{E_{e^\pm}(HI) \cdot E_{e^\pm}(LO)} \cr \cr
 \lambda_\pm &=& \sqrt{\frac{E_{e^\pm}(HI)}{E_{e^\pm}(LO)}}
 \end{eqnarray}
 where $\lambda = \gamma + \sqrt{\gamma^2 -1}$, and $\gamma = \lambda /2 + 1/(2 \lambda)$.
We can determine the relationship between the mass errors and the energy-edge
errors. To a very good approximation this follows from:
 \begin{eqnarray}
 \frac{\delta M_{\tilde{e}_{L,R}}}{M_{\tilde{e}_{L,R}}} &=&-~\frac{\delta\gamma_{\tilde{e}_{L,R}}}{\gamma_{\tilde{e}_{L,R}}} = \frac{\lambda_{\tilde{e}_{L,R}}^2 - 1}{\lambda_{\tilde{e}_{L,R}}^2 +1} 
 \bigg(  \frac{\delta \lambda_{\tilde{e}_{L,R}}}{\lambda_{\tilde{e}_{L,R}}}\bigg)
 \end{eqnarray}
This becomes to a good approximation 
\begin{eqnarray}
\frac{\delta M}{M} &=&
\frac{1}{\sqrt{2}}\frac{E_e(HI)~-~E_e(LO)}{E_e(HI)~+~E_e(LO)}\frac{\delta(E_e)}{E_e}
\end{eqnarray}
where we assume that $\delta(E_e)/E_e$ is the same at the low and high end of 
the energy spectrum.

The complete set of equations for determining the masses from the energy 
spectrum end-points is:
  \begin{eqnarray}
    2 \gamma_{\tilde{e}_R}  M_{\tilde{e}_R} \sqrt{s} &=& \sqrt{s}^2 + (M^2_{\tilde{e}_R} - M^2_{\tilde{e}_L}) \cr \cr
    2 \gamma_{\tilde{e}_L}  M_{\tilde{e}_L} \sqrt{s}  &=& \sqrt{s}^2 + (M^2_{\tilde{e}_L} - M^2_{\tilde{e}_R}) \cr \cr
   2E^*_{e,R} M_{\tilde{e}_R} &=& M^2_{\tilde{e}_R} - M^2_{\tilde{\chi}_1^0} \cr \cr 
   2E^*_{e,L} M_{\tilde{e}_L} &=& M^2_{\tilde{e}_L} - M^2_{\tilde{\chi}_1^0}  
  \end{eqnarray}
  The system, although quadratic, allows one to substitute the quadratic terms 
and reach a linear solution:
 \begin{eqnarray}
 M_R &=& \frac{1}{2} \sqrt{s} 
         \frac{\gamma_{\tilde{e}_L} \sqrt{s}   + 2 E^*_{e,L}}{\gamma_{\tilde{e}_L} \gamma_{\tilde{e}_R} \sqrt{s} +
               \gamma_{\tilde{e}_R}E^*_{e,L} + \gamma_{\tilde{e}_L} E^*_{e,R}}
  \cr \cr
 M_L &=& \frac{1}{2} \sqrt{s} 
         \frac{\gamma_{\tilde{e}_R} \sqrt{s}   + 2 E^*_{e,R}}{\gamma_{\tilde{e}_L} \gamma_{\tilde{e}_R} \sqrt{s} +
               \gamma_{\tilde{e}_L}E^*_{e,R} + \gamma_{\tilde{e}_R} E^*_{e,L}}
  \end{eqnarray}
The fit process that searches for the masses  $M_{\tilde{e}_L}$, $M_{\tilde{e}_R}$ and $M_{\tilde{\chi}_1^0}$ is carried out with MINUIT~\cite{bib:minu} that 
obtains the best solution to the set of equations above. 

\begin{figure}[t]
\epsfig{figure=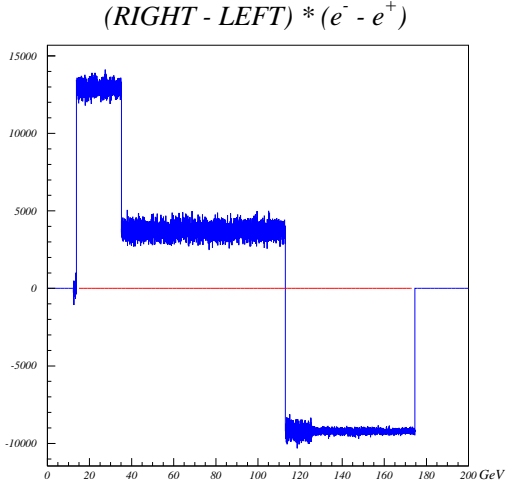,height=7.0cm}
 \caption{Energy distributions difference between $e^+$ and $e^-$ 
for only selectron decays. Using the difference of the difference, between 
$Right_{80\%}$ and $Left_{80\%}$ $e^-$ beam polarisations, the differences are 
enhanced and any detector asymmetries between charged tracks are effectively 
cancelled.}
 \label{fig:dde}
 \end{figure}

The properties of the above defined distribution are quite simple and robust. 
First, there are only 2 (well separated) ``boxes''  as seen in Fig.
\ref{fig:dde}; one positive, one negative, with an overlapping mid-region.
This means that the edges can be comfortably resolved, as none of them 
overlap in any energy range. Secondly, the edge-to-particle 
assignments are clear, the positive part of the distribution corresponding to 
$\tilde{e}^\pm_R$. Thirdly, since all SM backgrounds have the same energy 
distribution for the visible $e^+$ and $e^-$, their difference does not exhibit
any contribution from Standard Model backgrounds. Even if there is a detector 
asymmetry between positive and negative tracks, this is eliminated through the 
use of the polarised version of the above difference as described in eqn. 1 and
shown in Fig. 1.


It should be noted also that even though the backgrounds are reduced to the
level of their statistical fluctuations,  they can still be large and affect 
the signal. Hence removal of the background to first order by kinematical cuts
or strategically located detector vetoes is still useful. It has been shown~\cite{bib:fnax} that the $\gamma^*$$\gamma^*$ background can be 
suppressed to almost zero using these techniques. These cuts fail however when
$\tilde{e}^\pm_{L,R}$ and $\tilde{\chi}^0_1$ are close in mass, for this case 
a Very-Forward Veto device is crucial. 
The method proposed here greatly alleviates the kinematic severity of the cuts 
needed to control this background. 
\begin{figure}[t]
\epsfig{figure=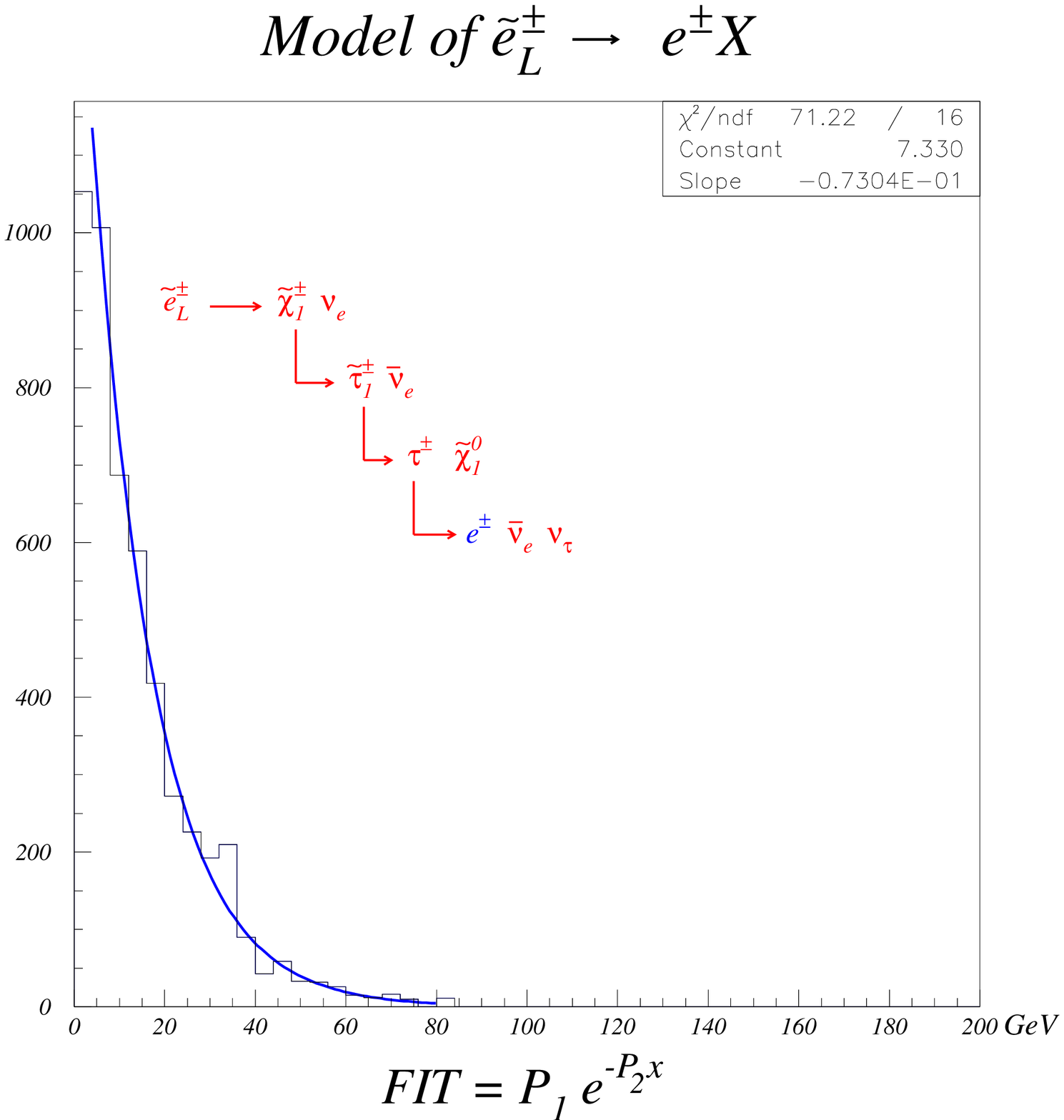,height=8.0cm}
 \caption{Parametrisation of the secondary leptonic decays of the $\tilde{\chi}^\pm_1$. The chosen model was an exponential, which requires one shape 
parameter.}
 \label{fig:ddx}
 \end{figure}

\begin{figure}[t]
\epsfig{figure=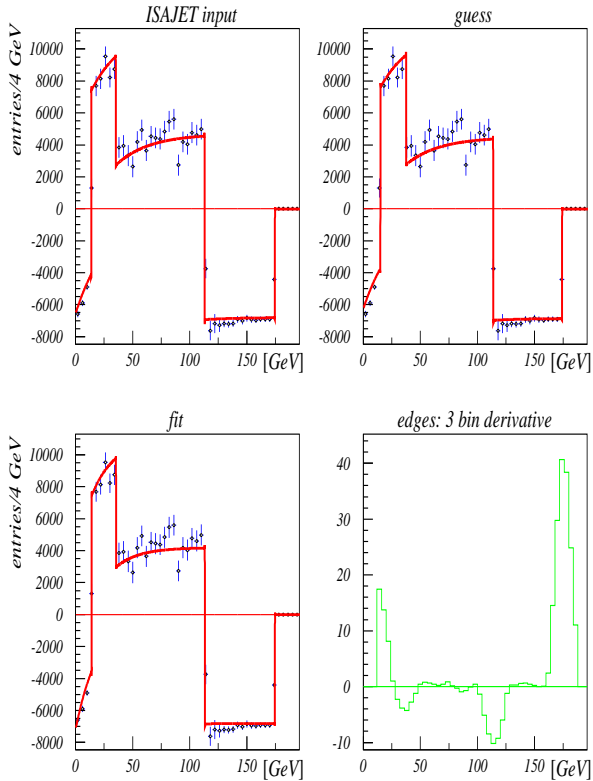,width=8.5cm,height=12.0cm}
\caption{Fit to the electron energy distribution including the $\tilde{\chi}^\pm_1$ decay mode of the $\tilde{e}_L$
based on an integrated luminosity of 2000 fb$^{-1}$ and a collision energy of 
500 GeV. The input masses (in GeV) are $M_{\tilde{e}_R}=142.96,
M_{\tilde{e}_L}=202.07, M_{\tilde{\chi}_1^0}=95.70$. The fit gives $M_{\tilde{e}_R}=143.12\pm 0.23, M_{\tilde{e}_L}=202.27\pm 0.11, M_{\tilde{\chi}_1^0}=95.72\pm 0.12$. }
 \label{fig:frl}
 \end{figure}

The procedure is complicated slightly by the fact that for a region of the SUSY
parameter space $\tilde{e}^\pm_{L}$ has two leptonic decay channels, one 
involving $\tilde{\chi}^\pm_1\nu_e$, and one $\tilde{\chi}^0_1e^\pm$. It is 
assumed that the secondary $\tilde{\chi}^\pm_1$ hadronic decays can be 
eliminated, however the leptonic decays with $\tilde{\chi}^\pm_1\nu_e$ have the
same signature, an $e^\pm$ in the final state as the preferred $\tilde{\chi}^0_1e^\pm$ decays and have to be accomodated in the fit. The difference 
distribution becomes in this case:
\begin{eqnarray}
\Delta(E&)&= Lumi \cdot
            (\sigma_{RL} - \sigma_{LR}) \cdot
            (f + f')  \,
	    \cdot \cr \cr
	     &\bigg[&R'_{box}(E) - \frac{f}{f + f'}
L'_{box}(E) - \frac{f'}{f + f'} X'_{box}(E)\bigg]
\end{eqnarray}
where $X'(E)$ is the energy distribution of the 
visible $e^\pm$ from the secondary $\tilde{\chi}^\pm_1$ decay, and $f$
and $f'$ are the branching fractions to $L'(E)$ and $X'(E)$ respectively.
The power of the method in its initial form is that this histogram has to be
normalised to zero, and hence any pedestal can be determined and subtracted. 
It also has 2 normalisation conditions related to the $R'(E)$ and $L'(E)$ boxes, that connect the edge positions with the box-heights. To first order there
are only 3 free parameters in the fit: $M_{\tilde{e}_L}$, $M_{\tilde{e}_R}$ and
$M_{\tilde{\chi}_1^0}$. This provides a very robust fit. In the present case, 
the $X'(E)$ distribution (figure \ref{fig:ddx}) has to be parametrised, and 
this adds an extra parameter to the fit.
The chosen model was an exponential shape for $X'(E)$, hence only one extra 
shape parameter. The general idea of the simple fit remains: a robust fit 
(figure \ref{fig:frl}) with few, tightly bound parameters. The fit yields also 
the $f/(f+f')$ and $f'/(f+f')$ relative leptonic branching fractions for the 
$\tilde{\chi}^\pm_1\nu_e$ and $\tilde{\chi}^0_1e$. The $X'(E)$ shape parameter 
gives general information about the $\tilde{\chi}^\pm_1$ and $\tilde{\tau}^\pm_1$ masses.

\begin{figure}[t]
\epsfig{figure=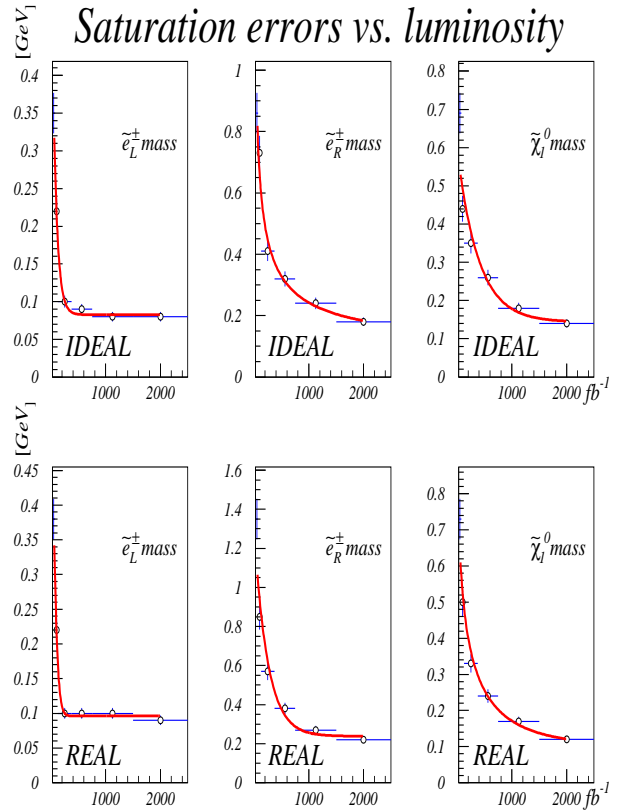,width=9.0cm,height=12.0cm}
\caption{Mass determination errors {\em versus} luminosity, for the
prototype distribution that includes only the $\tilde{\chi}^0_1e$ decay and for
the realistic one including also the $\tilde{\chi}^\pm_1\nu_e$ decay. An 
indication of saturation with luminosity is seen.}
 \label{fig:vsl}
 \end{figure}

This analysis is carried out for the MSSM parameters m$_0$=100 GeV, m$_{1/2}$=250 GeV, A$_0$=0, tan($\beta$)=10, $\mu$=positive. The resultant masses (in GeV)
of the supersymmetric particles in our study is m$_{\tilde{e}_R}$=143 , m$_{\tilde{e}_L}$=202, m$_{\tilde{\chi}_1^0}$=96, m$_{\tilde{\chi}_1^\pm}$=174, and 
m$_{\tilde{\tau}_1^\pm}$=135. The errors in the mass determination {\em versus}
luminosity are shown in 
figure \ref{fig:vsl} for the ``idealistic" case with no $\tilde{\chi}^\pm_1\nu_e$ channel, as well as for the ``realistic" case including the $\tilde{\chi}^\pm_1\nu_e$. An indication of saturation of the mass-errors with luminosity is 
shown by the fit. The luminosity at which the error reaches 80\% of its 
limiting value gives a range of 500 - 1000 $fb^{-1}$ as the saturation limit.
It can be seen that the largest difference in mass-errors between the ``ideal''
and  ``real" case is for $M_{\tilde{e}_R}$ due to its corresponding energy-``box'' neighboring the perturbing $X'(E)$ distribution. 

The method is very robust and stable, its results being not only the masses
involved, but also possibly the relative leptonic branching ratios, and some 
tangential information about the $\tilde{\chi}^\pm_1$ and $\tilde{\tau}^\pm_1$ 
masses. This last aspect is currently under study. The resolution obtained is 
comparable to that obtained with production threshold studies~\cite{bib:Martyn,bib:Peskin}. A study of the mass 
determination errors {\em versus} luminosity indicates a possible saturation 
of the errors from the $1/\sqrt{N}$ law, occuring when the luminosity is of the
order of 500 - 1000 $fb^{-1}$.

We would like to thank our colleagues that participated in the SUSY 
Snowmass-2001 studies, specially Jonathan Feng, Paul Grannis, and Robert Kahn, 
for their comments on this work. This work was carried out with the support of 
the DOE under grant DE-FG03-95ER40894.


\begin{thebibliography}{99}

\bibitem{bib:tsukamoto}
Toshifumi Tsukamoto, Keisuke Fujii, Hitoshi Murayama, Masahiro Yamaguchi, and 
Yasuhiro Okada, ``Precision Study of Supersymmetry at Future e$^+$e$^-$ Colliders,'' Phys. Rev D 51, 3153 (1995). 

\bibitem{bib:Feng1}
J.~L.~Feng and D.~E.~Finnell,
``Squark mass determination at the next generation of linear e+ e-
colliders,''
Phys.\ Rev.\ D {\bf 49}, 2369 (1994)
[hep-ph/9310211].

\bibitem{bib:Feng2}
J.~L.~Feng, M.~E.~Peskin, H.~Murayama and X.~Tata,
``Testing supersymmetry at the next linear collider,''
Phys.\ Rev.\ D {\bf 52}, 1418 (1995)
[hep-ph/9502260].

\bibitem{bib:Martyn}
``Supersymmetry, Chapter 3 of TESLA TDR'', DESY-2001-011, ECFA-2001-209.

\bibitem{bib:Baer}
H.~Baer, R.~Munroe and X.~Tata,
``Supersymmetry studies at future linear e+ e- colliders,''
Phys.\ Rev.\ {\bf D 54}, 6735 (1996)
[hep-ph/9606325].

\bibitem{bib:Colorado}
The various studies by the Colorado group are discussed in http://hep-www.colorado.edu/SUSY.

\bibitem{bib:minu}
MINUIT v94.1 minimisation program, F.~James and F.~Roos,
CERN (1967).

\bibitem{bib:fnax}
SUSY-$\tilde{e}^\pm$ Studies, The Colorado Group,
LCWS-2000 Proceedings, Fermilab (2000).

\bibitem{bib:Peskin}
J.~L.~Feng and M.~E.~Peskin,
``Selectron studies at e- e- and e+ e- colliders,''
Phys.\ Rev.\ D {\bf 64}, 115002 (2001)
[hep-ph/0105100].


\end{thebibliography}
\end{document}